\documentclass[preprint]{revtex4}
\usepackage{amsfonts}
\usepackage{amssymb}
\usepackage{amsmath}
\usepackage{graphicx}
\usepackage{array}
\usepackage{dcolumn}
\usepackage{color}
\newcommand{\WA}{$H_2O$}

\newcommand{\pbas}{\vert j_1k_1m_1j_2k_2m_2\rangle }
\newcommand{\pbasd}{\langle j_1k_1m_1j_2k_2m_2\vert }
\newcommand{\symtb}{\vert j \bar{k} \bar{m} \rangle}
\newcommand{\symt}{\vert jkm \rangle}
\newcommand{\wigs}{D^{j*}_{mk}(\alpha, \beta, \gamma)}
\newcommand{\wig}{D^{j}_{mk}(\alpha, \beta, \gamma)}
\newcommand{\cn}[1]{$(H_2O)_{#1}$}

\begin{document}
\title{Monomer Basis Representation Method For Calculating The Spectra Of 
         Molecular Clusters II. Application  
         To Water Dimer.}
\author{Mahir E. Ocak}
\email{meocak@alumni.uchicago.edu}
\affiliation{Ya\c{s}amkent Mahallesi, Yonca Sitesi 13/B Daire No:5, \c{C}ayyolu,
		Ankara, Turkey
            }
\date{\today}
\begin{abstract}
The  Monomer Basis Representation (MBR) method 
developed in the first paper 
is applied to water dimer in order to illustrate its 
application and to show its validity.
The calculations are done by using the SAPT-5st potential surface.
Monomers are treated as rigid bodies. Radial coordinate is separated
from the angular coordinates adiabatically. MBR method is used for solving
the five dimensional angular problem. Then, the results of the angular
calculations are fit to a Morse function to find the potential surface
for the radial motion. The results
show that the method works efficiently and
accurately.
\end{abstract}
%\pacs{34.20.Gj, 36.40.-c, 36.40.Mr}
%\keywords{Monomer Basis Representation (MBR) Method, water clusters,
%             water dimer, potential surfaces}
\maketitle

\section{\label{sec:intro}Introduction}

Water clusters have been the subject
of intense research in the literature,
since they are important for understanding hydrogen bonding, for interpreting
the many special futures of the structure, dynamics and energetics of
bulk solid and liquid water. Besides hydrogen bonding to water molecules is
of general importance in many biological, chemical and physical systems.

Theoretical studies of small water clusters shows that many-body terms
are very important.
For water, it is known that many-body terms
 may account for up to
$25\%$ of the interaction energy of the bulk water
\cite{hermansson94}.
It has been found that
\cite{wormer99}:
the pair interaction energy represents $83-86\%$ of the total interaction energy
of \cn{3}. For \cn{4} the same percentage amounts to $\approx 75\%$,
for \cn{5} only to $\approx 68\%$. It is evident that many-body terms are
far from negligible. It has also been found that the major
part of the many-body cooperative effect comes from three-body terms. The
four-body terms represent $2\%$ of the total interaction energy of the \cn{4},
 and less than $4\%$ of the total interaction energy of the \cn{5}.
Therefore, characterization of three body terms is a major step for the
development of an accurate potential surface
 for bulk liquid and solid water.

Before starting to work on three-body terms, first it is necessary to
have an accurate description of the two body terms in the potential surface.
 For that purpose,
 water dimer has been studied extensively both
experimentally
\cite{dyke77a,dyke80,Coudert87,huang,huang88,Busarow89,dyke89,fraser89,fraser91,zwart91,keutsch3,braly,braly2,loeser93}
 and theoretically
 \cite{clary2,Clary95,clary95-2,leforestier1,leforest99,jcl155,avoirdpes00,avoirdpes00a}, and many potential surfaces has been developed
\cite{MCY,watts82,jorgensen81,jorgensen82,jorgensen83,berendsen87,cieplak90,berweger95,millot98}.

Among the theoretical studies of water dimer, Clary and co-workers were
the first ones to perform six dimensional calculations. Firstly, Althorpe
and Clary studied water dimer by separating the stretching coordinate
form the angular coordinates adiabatically \cite{clary2}. This adiabatic
approximation was justified later with a more exact treatment \cite{jcl155}.
In the calculations, 
Wigner rotation functions were used as angular basis for each monomer, and
the molecular symmetry of water dimer was fully exploited. 
Later, the same authors developed a new method named 
Discrete Variable Representation
Iterative Secular Equation (DVR-ISE) which is an extension of the
diagonalization truncation method by using the variation-perturbation 
theory. The method utilizes the iterative secular equation (ISE) method
\cite{Slee93}.
In the calculations, functional bases were used for some of the angles and
DVR bases were used for the other angles. Later, Gregory and Clary studied 
water dimer \cite{clary95-2} with Diffusion Monte Carlo (DMC) method
\cite{Sun90,Quack91}. Leforestier and co-workers were the first ones to do
fully coupled six dimensional calculations with basis sets. They published 
two papers \cite{leforestier1,leforest99}. The first one included some 
erroneous data because of a mistake in the moments of inertia of water 
monomers which were being corrected in the second one.
 The calculations were done with a coupled product basis of Wigner 
rotation functions 
by using the pseudospectral split Hamiltonian (PSSH) formalism in which
the kinetic energy terms are evaluated in the coupled product basis
of Wigner rotation functions
and the potential energy is evaluated in the grid basis. Later, Chen and
Light also made fully coupled six dimensional calculations of water dimer
\cite{jcl155}
by using the sequential diagonalization truncation method in which angular and
the radial parts of the Hamiltonian were diagonalized successively. In these
calculations, authors used the same decoupled product basis of Wigner 
rotation functions with the 
first calculations of Althorpe and Clary \cite{clary2}, and the symmetry 
of water dimer was exploited fully in calculations. Results showed that
the adiabatic approximation of the Althorpe and Clary is successful for
predicting the tunneling splittings. Finally, van der Avoird and co-workers
developed a new potential surface called SAPT-5s \cite{avoirdpes00}
by using the Symmetry Adapted Perturbation Theory (SAPT) 
\cite{avoird95,avoird95e,wormer99}. This potential 
surface was tuned for predicting the vibration-rotation-tunneling 
levels of water dimer which led to the development of a new 
potential surface called SAPT-5st \cite{avoirdpes00a}.
 The tuned  potential surface 
describes the experimental data with near spectroscopic accuracy
\cite{avoirdpes00,avoirdpes00a,avoird-prl}. In calculations, 
first an  equally spaced grid was taken for the stretching coordinate 
and the angular Hamiltonian was solved at $49$ different grid points by 
using a coupled product basis of Wigner rotation functions.
 Then, a three point contracted 
DVR \cite{Echave92} basis were used for the stretching  coordinate.  

 Since some experimental data
about water trimer is available \cite{trimerrev,pugliano92,keutsch1,liu5,viant,keutsch2,viant99},
 many researchers started to
work on water trimer
 \cite{trimerI:96,trimerII:96,schutz93,wyatt98,janssen99,szalewicz2003,bacic-cpc,xantheas93-2,schutz95,brown2}. However, since
the size of the water trimer problem is too big to handle quantum mechanically,
all the theoretical studies done up to now are based on very reduced
dimensionality models, even for rigid water molecules. The method developed
in the first paper may make it possible to study clusters as big as water trimer
with more realistic models as it will be seen. 

In the following section, the MBR method developed in the first paper 
will be used for calculating the vibration-rotation-tunneling (VRT) 
spectra of water dimer in order to illustrate the application of the 
method, and to show its validity.  In calculations, 
adiabatic approximation of Althorpe and Clary, PSSH formalism of Leforestier
and co-workers 
and the potential surface of Groenenboom {\it et. al.} will 
be used. The main difference of the calculations here from the previous
calculations is the generation of optimized bases for each monomer 
in the cluster by using the MBR method. It will be seen that 
the use of the MBR method leads to successful results with a basis which has 
a much smaller size
 than any of the bases used in previous studies of water dimer.

\section{Structure of Water Dimer}

Water clusters are highly nonrigid, even with rigid monomers.
 Their potential surface
contains more than one global minimum which are equivalent by symmetry.
The potential barriers between these minima are not high, so that
the molecule can tunnel through these barriers which causes tunneling
splittings in the vibration-rotation-tunneling (VRT)
 spectra of the molecules. Calculation of these splittings is a good
test for the accuracy of the available potential surfaces. Since the wave 
functions corresponding to the splitting levels has considerable amplitude
in the region where the tunneling occurs, a good prediction of the tunneling 
splittings shows that the potential surface used in the calculations is 
accurate not only around the minimum (which is usually the case when there
is no tunneling) but also in the tunneling region.

The structure of water dimer was first determined by Dyke and co-workers
via rotation spectra 
\cite{dyke77a,dyke80}. By examining the experimental data, Dyke 
realized the presence of tunneling splittings and made a group theoretical
classification of the tunneling-rotational levels 
\cite{dyke77}
 by using Permutation Inversion (PI)
group theory \cite{bunker98,Bunker2005}.
It was shown that, the equilibrium structure of water dimer has 
a plane of symmetry and a nonlinear hydrogen bond. The equilibrium
structure of water dimer is depicted roughly in figure \ref{fig:bffdimer}.
 The $O-O$
distance reflecting the length of the H-bond is 2.952 \AA, and the dissociation
energy of the  hydrogen
bond, $D_e$ is 3.09 kcal/mole \cite{keutsch1}. 

By including all feasible permutation inversion operations it is possible 
to generate 16 different configurations. Due to the presence of a plane of
symmetry in the equilibrium structure,
 there is twofold structural degeneracy and 
only eight of these structures are non-superimposeable. 
There exist three distinct tunneling motions that 
rearrange the H-bond network on time scales ranging from 1 {$\mu$}s to 1 ps
\cite{saykally-s99}. The tunneling motions connect eight degenerate minima
on the intermolecular potential surface (IPS). These tunneling motions
are acceptor switching (AS), 
interchange tunneling (I) and  bifurcation
tunneling (B). 

Acceptor switching 
has the smallest barrier for the tunneling. In that motion
the protons of the acceptor monomer exchange their positions. This tunneling
does not break the hydrogen bond. Its causes splitting of each ro-vibrational
level into two levels.

Interchange tunneling 
interchanges the roles of the acceptor and donor 
monomers. This motion has the second lowest barrier and it causes 
further splitting of the ro-vibrational levels.

Bifurcation tunneling 
has the highest barrier. Since acceptor switching 
and interchange tunneling, together with the inversion of the dimer,
 resolves all the degeneracy of the water dimer. 
The effect of the bifurcation tunneling is not to split the energy levels,
but just to shift them. This is a result of the fact that water dimer
has a plane of symmetry, which causes two fold structural degeneracy of 
each ro-vibrational level. Acceptor switching, interchange tunneling
and the inversion of the dimer already connects eight degenerate minima
to each other. 

%\begin{figure}[htbp]
%\caption[Equilibrium structure of water dimer.]
%{Equilibrium structure of water dimer as it is determined
%by calculations on the VRT(ASP-W)-II potential surface
%\cite{keutsch1}. The hydrogen bond deviates from linearity by 
% $2.3^{\circ}$, the O-O distance is 2.952 \AA, and the bond
%strength, $D_0$, is 3.40 kcal/mole. This figure is taken from 
%a paper by Keutsch and Saykally 
%\cite{keutsch1}.}
%\label{dimer}
%\begin{center}
%\includegraphics[scale=0.8]{dimer3.eps}
%\end{center}
%\end{figure}

%\begin{figure*}[htbp]
%\caption[Tunneling motions of water dimer.]
%{Water dimer exhibits three distinct tunneling motions.
%Acceptor switching (AS) exchanges the two protons in the hydrogen bond
%acceptor monomer. Interchange tunneling (I) exchanges the roles of the
%hydrogen bond donating and accepting monomers. In the bifurcation tunneling
%(B), the hydrogen bond donor exchanges its protons. This figure is taken
%from a paper by Keutsch and Saykally \cite{keutsch1}.}
%\label{dimer-tun}
%\begin{center}
%\includegraphics[angle=270,scale=0.5]{dimer-tun.eps}
%\end{center}
%\end{figure*}

\begin{figure}
\caption[Correlation diagram for the rotation-tunneling states of water dimer
for $J=0$.]{Correlation diagram for the rotation-tunneling states of
\cn{2} for $J=0$. In the figure AS, I and B refers to acceptor
switching, interchange tunneling and bifurcation tunneling respectively. Levels
are labeled with the irreducible representations of the $G_{16}$ PI group.}
\label{dimer-split} 
\begin{center}
\input{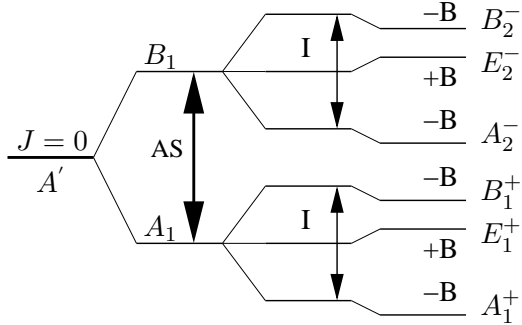}
\end{center}
\end{figure}

The splittings for $J=0$ rotational level of water dimer are 
shown in figure \ref{dimer-split}.
 Each energy level is labeled with 
the irreducible representations of the $G_{16}$ PI group which is 
the molecular symmetry group of water dimer.  

\section{Hamiltonian And The Outline of Calculation Strategy}

The full water dimer problem is too big to handle quantum mechanically. 
Therefore, it is necessary to model the problem. In weakly bound clusters,
inter-molecular and intra-molecular degrees of freedom have frequencies 
which differ by at least one order of magnitude. As a result of that
inter-molecular and intra-molecular degrees of freedom can be separated
adiabatically. Thus, while treating the inter-molecular degrees of freedom,
the monomers can be considered as rigid bodies. 

The Hamiltonian for the inter-molecular motion of a nonrigid system 
consisting of two rigid polyatomic fragments can be written as \cite{brocks83} 
\begin{equation}
\hat{H}= - \frac{1}{2\mu R} \frac{{\partial}^2}{\partial R^2} R + \hat{K_1}
+ \hat{K_2} + \hat{K_{12}} + \hat{V}.
\label{hamiltonian}
\end{equation}
In the equation above,  $R$ is the 
distance between the centers of mass of the monomers.
$\mu$ is the reduced mass of the dimer given by 
\begin{equation}
\frac{1}{\mu}=\frac{1}{M_1}+\frac{1}{M_2},
\end{equation}
where $M_i$ is the total mass of the monomers.

In the Hamiltonian $\hat{K_i}$ is the kinetic energy operator of the monomer, 
which can be expressed in the body fixed frame of the monomer as 
\begin{equation}
\hat{K_i}=A \hat{j_{ix}^2} + B \hat{j_{iy}^2} + C \hat{j_{iz}^2},
\end{equation}
in terms of the angular momentum operators around the body fixed axis of the 
monomer, 
which is the molecular symmetry axis in this case. For calculations,
the $z$ axis is defined to be the bisector of the $HOH$ angle.  The plane
of the molecule is defined to be the $xz$ plane. Rotational constants 
$A,B,C$ are given in terms of the moments of inertia as
\begin{equation}
A=\frac{1}{2I_x}, \; \; B=\frac{1}{2I_y}, \;  \; C=\frac{1}{2I_z},
\end{equation}
where $I_x, I_y, I_z$ are the moments of inertia of the monomers around
$x,y,z$ axes respectively. Since the monomers are considered to be rigid
$A,B$ and $C$ are constants. 

In equation (\ref{hamiltonian}), $\hat{K_{12}}$
defines  the kinetic energy operator corresponding to the 
end-over-end rotation of the dimer. It is given by
\begin{equation}
\hat{K_{12}}=\frac{1}{2\mu R^2}(\hat{J^2}+\hat{j^2}-2\hat{J}.\hat{j}).
\end{equation}
In the equation above $\hat{J}$ is the total angular momentum of the
system given by 
\begin{equation}
\hat{J}=\hat{j_1}+\hat{j_2}+\hat{L},
\end{equation}
where $\hat{j_i}$'s are the monomer angular momentum operators, 
and $\hat{L}$
is the angular momentum operator for the end-over-end rotation of the dimer.
$\hat{j}$ is defined as $\hat{j}=\hat{j_1}+\hat{j_2}$. 

\begin{figure}
\caption{\label{fig:bffdimer} Body fixed frame of water dimer.}
\begin{center}
\input{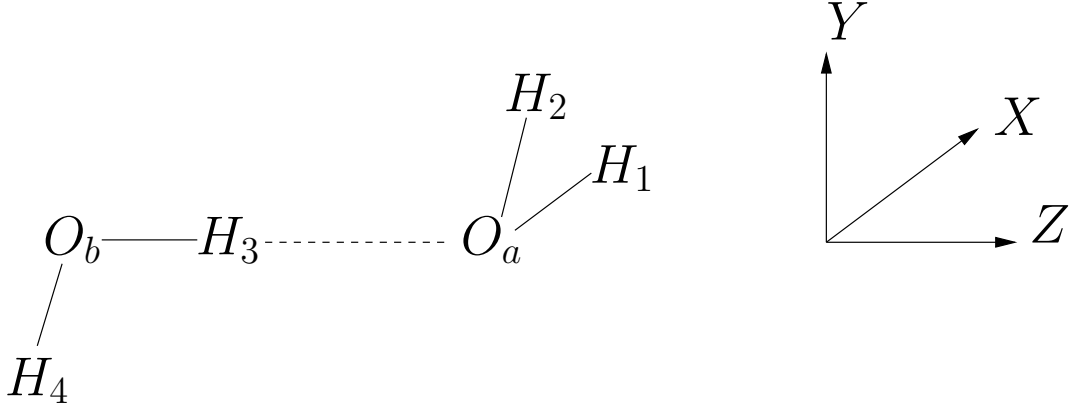}
\end{center}
\end{figure}

In order to evaluate the $\hat{K_{12}}$ term, it is necessary to express
all the angular momentum operators in a common reference system. This 
reference system is the body fixed coordinate system of the dimer, of course.
The $z$ axis of the body fixed  frame of the dimer 
is defined along the line joining the center of
mass of the monomers  and the $y$ axis of the body fixed frame of the dimer
is defined to be along
the bisector of the $HOH$ angle of the acceptor monomer in the equilibrium 
configuration; see figure \ref{fig:bffdimer}.
  In this reference frame, $\hat{K_{12}}$ is expressed as 
\begin{equation}
\hat{K_{12}}= \frac{1}{2\mu R^2}(\hat{j}_1^2+ \hat{j}_2^2 + 
2 \hat{j}_{1z}^{'} \hat{j}_{2z}^{'}+ \hat{j}_{1+}^{'} \hat{j}_{2-}^{'}
+\hat{j}_{2+}^{'} \hat{j}_{1-}^{'}
-2\hat{J}_z^{'}\hat{j}_z^{'}-\hat{J}_+^{'}\hat{j}_-^{'}- 
\hat{J}_-^{'}\hat{j}_+^{'}).
\end{equation}
For $J=0$, this equation reduces to 
\begin{equation}
K_{12}= \frac{1}{2\mu R^2}(\hat{j}_1^2+ \hat{j}_2^2 +
2 \hat{j}_{1z}^{'} \hat{j}_{2z}^{'}+ \hat{j}_{1+}^{'} \hat{j}_{2-}^{'}
+\hat{j}_{1+}^{'} \hat{j}_{2-}^{'}).
\end{equation}
In the equations above $'$ denotes that the operator refers to the 
body fixed frame of the dimer but not the body fixed frame of the monomers.

In order to solve the eigenvalue problem, first the stretching coordinate 
will be separated from the angular coordinates adiabatically. This is 
first done by Althorpe and Clary \cite{clary2}, and led to successful
results. 

First, the angular Hamiltonian is written in the form 
\begin{equation}
\hat{H}_{ang}=\hat{h_1^0}+\hat{h_2^0}+\hat{\Delta K} + \hat{\Delta V}
\label{partition}
\end{equation}
where the model Hamiltonians, $\hat{h_i^0}$, for the three dimensional
monomer problems are given by
\begin{equation}
	\hat{h_i^0}=\hat{K_i}+\hat{V_i^0}
\end{equation} 
in which $\hat{K_i}$ is the monomer's kinetic energy operator in equation
(\ref{hamiltonian}), and $\hat{V_i^0}$ is
 the model potential energy surface for the 
three dimensional problem, which is the rotation of the monomers. 

By comparing equation (\ref{partition}) with the Hamiltonian, equation 
\ref{hamiltonian}, it is easily seen that $\Delta K= \hat{K_{12}}$ 
and $\Delta V$ terms can be identified as 
\begin{equation}
	\hat{\Delta V}= \hat{V} - \hat{V_1^0} - \hat{V_2^0}.
\end{equation}

The results of the angular calculations will be used to find an effective
potential surface for the radial coordinate so that the total Hamiltonian
can be written as 
\begin{equation}
	\hat{H}= - \frac{1}{2 \mu R} \frac{\partial^2}{\partial R^2}
			+ V_{eff}(R).
\end{equation} 

Details of monomer calculations, five dimensional angular calculations
and radial calculations 
will be given in sections \ref{ssec:moncal}, \ref{ssec:fdcal} and \ref{sec:radcal},
 respectively.

\section{Generation of A Monomer Basis}
\label{ssec:moncal}

\subsection{Model Potential Surface}
For solving the monomer problem, it is first necessary to chose a model
potential surface. The choice here will be the external field which 
is generated by the other monomer. 

When one wants to solve eigenstates of one monomer in the field of the 
other one, a natural question arises. Since
the monomers in water dimer are not in the same environment their
calculations will result in  different eigenstates. Since the dimer includes
both a donor and an acceptor, the results of neither calculation will 
generate a sufficient basis for solving the five  dimensional angular problem.
In order to handle this donor acceptor asymmetry of water dimer, 
the Energy Selected Basis (ESB) 
method will
 be used \cite{hsl1,hsl2}.

In the ESB method, 
one generates the model potential surface as 
a marginal potential in which the potential energy  
at a point is the minimum value of the  potential energy with respect
to all other coordinates, i.e.
\begin{equation}
	\hat{V^0}(q_k)=\hat{V^0}(q_k,q_{k'}^{min}).
\end{equation}

Since the coordinates of the monomer which causes the external field is
varied to find the minimum potential, the marginal potential will sample
both the donor and the acceptor configurations. Therefore, 
the monomer calculation should result in a sufficient basis for both the donor
and the acceptor monomers.

For solving water dimer problem with energy selected basis method,
it is necessary to generate a three dimensional marginal potential. Labeling the
monomers as $i$ and $j$, the model potential for monomer $i$ in the field 
of monomer $j$ will be  
\begin{equation}
\hat{V^0_i}(\zeta_i)= \hat{V}(\zeta_i,\zeta_j^{min};R)
\end{equation}
where $\zeta_i=(\alpha_i,\beta_i,\gamma_i)$ and $\zeta_j=(\alpha_j,\beta_j,\gamma_j)$
 refer to the Euler angles of monomers which describe the 
orientation of the body fixed frame of the monomers with respect to the
body fixed frame of the dimer. 

\subsection{Primitive Bases}

In order to solve the problem two primitive bases will be used, one 
functional basis and one grid basis. Functional basis being small and 
compact will be the primary basis. Grid basis will be used for the evaluation 
of potential energy.
Primitive functional basis  will be 
 the symmetric top basis $\vert jkm \rangle$. In this basis, $j$ is the
total angular momentum, $k$ is the projection of angular momentum onto the
body fixed $z$ axis of the monomer and $m$ is the projection of 
angular momentum on to the body fixed $z$ axis of the dimer.
Symmetric top basis functions 
 can be expressed in terms of Wigner rotation functions as 
\begin{equation}
 \symt =  \frac{1}{2 \pi}\sqrt{\frac{2j+1}{2}} \wigs ,
\label{pbas}
\end{equation}
where $\wig$ is the the Wigner rotation function, expressed in terms of Euler 
angles $\alpha$, $\beta$ and $\gamma$. Its functional form is given by 

\begin{equation}
\wig= e^{-im\alpha} d^j_{mk}(\cos \beta) e^{-ik\gamma}.
\label{wigfun}
\end{equation} 
Exact functional form and the symmetry properties of the $d^j_{mk}(\cos \beta)$
 function
can be found elsewhere \cite{rose}.

Following Light \cite{jcl91,jcl117,jcl160}, the representation of the Hamiltonian in the symmetric top  
basis can be called
 Finite Basis Representation (FBR), and the representation
of Hamiltonian in the grid basis can be called 
Discrete Variable Representation
(DVR).  
The application of DVR methodology to non-product bases is
discussed by Corey {\it et. al.} \cite{lemoine97,corey93} for the case of
spherical harmonics, and later it is 
applied to the case of symmetric top basis 
\cite{leforest,leforestier1,leforest99}. The deviation 
from Light's original formulation in the case of coupled bases is that 
one no longer seeks a unitary 
transformation between two bases. Therefore, two representations 
are no longer equivalent. Since the FBR basis is more 
compact it is the primary basis used in the calculations.  Use of the DVR basis
provides a simple way for the evaluation of potential energy matrix elements,
since the potential energy matrix is diagonal in the DVR basis, and the 
matrix elements are given by the value of the potential energy at the 
corresponding grid point. 

\subsection{Symmetry Adaptation of Basis Functions} 
\label{ssec:monsym}
As discussed in the first paper, the monomer problem
should be symmetry adapted to the direct product group of the pure permutation
group of the monomer, and the inversion subgroup of the dimer. 
If one labels the hydrogen atoms as $1$ and $2$, then the
 pure permutation group of water monomers is the  
group $G_2=\{E,(12)\}$. The inversion subgroup is 
as usual $\varepsilon= \{E, E^*\}$.
In this case, the direct product group becomes isomorphic to the molecular 
symmetry group of free water monomers, which is $C_{2v}(M)$. 
However, 
the $E^*$ operation here refers to the inversion of the
full dimer system, and not to the inversion of free water monomers.
Character table of the group $C_{2v}(M)$ is given in table 
\ref{tab:chartabc2vm}.

\begin{table}[p!tb]
\caption{\label{tab:chartabc2vm}Character table of the $C_{2v}(M)$
permutation inversion group. This group is the direct product of
the groups $G_2$, whose character table is given in 
table \ref{tab:chartabg2}, 
and $\varepsilon$, whose character table is given in table
 \ref{tab:chartabeps}. In the table, $\Gamma = x \otimes y$ means that
the irreducible representation $\Gamma$ of the group $C_{2v}(M)$ is obtained
as direct product of the irreducible representation $x$ of the group $G_2$ and
the irreducible representation $y$ of the group $\varepsilon$.}
\begin{center}
\begin{tabular}{crrrr}
\hline \hline
$C_{2v}(M)=G_2 \otimes \varepsilon$ & $E$ & $(12)$& $E^*$ & $(12)^*$ \\
\hline
$A_1=A\otimes G$ & 1 & 1 & 1 & 1 \\
$A_2=A\otimes U$ & 1 & 1 & -1 & -1 \\
$B_1=B\otimes U$ & 1 & -1 & -1 & 1 \\
$B_2=B\otimes G$ & 1 & -1 & 1 &  -1 \\
\hline \hline
\end{tabular}
\end{center}
\end{table}

\begin{table}[p!tb]
\caption{\label{tab:chartabg2}Character table of the $G_2$
permutation group.} 
\begin{center}
\begin{tabular}{crr}
\hline \hline
$G_2$ & $E$ & $(12)$ \\
\hline
A & 1 &1 \\
B & 1 & -1 \\
\hline \hline
\end{tabular}
\end{center}
\end{table}

\begin{table}[p!tb]
\caption{\label{tab:chartabeps}Character table of the
   inversion group $\varepsilon$.}
 \begin{center}
\begin{tabular}{crr}
\hline \hline
$\varepsilon$ & $E$ & $E^*$ \\
\hline
$G$ & 1 &1 \\
$U$ & 1 & -1 \\
\hline
\end{tabular}
\end{center}
\end{table}

\begin{figure}
\caption[Orientation of the body fixed frame of water monomers.]{Orientation of the body fixed frame of  water monomers.
 Molecule is
in the $xz$ plane. Origin of the axes is the center of mass of 
the water molecule.} 
\label{monff}
\begin{center}
\input{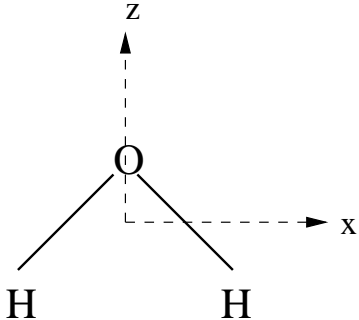}
\end{center}
\end{figure}

In order to make the symmetry adaptation of the basis functions, it 
is necessary to find the effects of the 
permutation inversion operations
to Euler angles and to the basis functions. A simple way of finding 
the transformation properties of Euler angles 
is given in appendix 
\ref{app:transeu}. By using the transformation properties  given in 
that appendix and by defining the body fixed frame of the monomers as 
in figure \ref{monff}  
effect of the permutation operations to Euler angles can be found. 
The results are given in table \ref{tab:monetr}. By using these 
transformations and the symmetry properties of the Wigner rotation 
functions, 
the effects of the permutation inversion operations to the 
symmetric top basis 
functions can also be found, the results are given in 
 table  \ref{tab:monetr}.

By using the projection operators of the group $C_{2v}(M)$, whose character
table is given in table \ref{tab:chartabc2vm}, symmetry adaptation 
of the basis functions can be done. The symmetry adapted basis functions
will be in the form of
\begin{equation}
   \vert jkm;p \rangle = N_p \left( \symt + (-1)^p \symtb \right),
\label{safbr}
\end{equation}
with the normalization constant given by,
\begin{equation}
N_p=\frac{1}{2(1+\delta_{k0}\delta_{m0})}.
\end{equation}

In equation (\ref{safbr}), $p$ is either $0$ or $1$, and $k$ is 
either even or odd depending on 
the symmetry.  Table \ref{parmonbas} gives the values of the parameter $p$
for each symmetry level.

\begin{table}
\caption{Transformation properties of Euler angles under the effect of the 
symmetry
operations of  the $C_{2v}(M)$ molecular symmetry group and the 
effect of the permutation inversion operations to symmetric top
basis functions.}
\label{tab:monetr}
\begin{center}
\begin{tabular}{ccccc}
\hline \hline 
$E$    & $\alpha$ & $\beta$ & $\gamma$ & $\symt$  \\ \hline
$(12)$ & $\alpha$ & $\beta$ & $\pi+\gamma$ & $(-1)^k \symt$\\
$E^*$  & $\pi-\alpha$ & $\beta$ & $-\gamma$ & $(-1)^k \symtb$ \\
$(12)^*$ & $\pi-\alpha$ & $\beta$ & $\pi-\gamma$ & $\symtb$ \\
\hline \hline
\end{tabular}
\end{center}
\end{table}

\begin{table}
\caption[Partitioning of the symmetry adapted functional basis of a monomer for 
water dimer calculations.]{Partitioning of the symmetry adapted monomer
basis for the monomer calculations. Form of the 
basis functions is given in equation (\ref{safbr}).}
\label{parmonbas}
\begin{center}
\begin{tabular}{c|c|c}
Representation & $p$ & $k$ \\
\hline
$A_1$ & $0$ & even\\
$A_2$ & $1$ & even\\
$B_1$ & $0$ & odd\\
$B_2$ & $1$ & odd
\end{tabular}
\end{center}
\end{table} 

Symmetry adaptation of the grid basis can also be done. Construction
of a grid basis for Wigner rotation functions 
is discussed by 
Leforestier \cite{leforest,leforestier1,leforest99}. 
According to his prescription, the grid basis that corresponds to an 
FBR 
basis of symmetric top basis 
is a uniform grid
for the angles $\alpha$ and $\gamma$ whose range is $(0,2\pi)$,
and  Gauss-Legendre 
quadrature points for $\cos \beta$ whose range is $(-1,1)$.
 This choice of grid basis corresponds to a direct product 
basis of plane wave DVR bases in angles 
$\alpha$ and $\gamma$ and a Gauss-Legendre 
DVR basis in $\cos \beta$. 

The construction of symmetry adapted DVR functions 
is 
discussed by Light and 
Carrington \cite{jcl160}. There are two approaches. First approach is to 
find the symmetry adapted DVR functions by diagonalizing a symmetric function
of the coordinate in an FBR basis which has the desired symmetry properties
\cite{jcl112}. Second approach is to find the DVR basis functions
by first using the usual procedure  and then to obtain the symmetry adapted
DVR functions as linear combinations of primitive DVR functions
 by using the well known
projection operators technique. This method is first used by Carrington
\cite{Wei:94,McN:93}. If the application of symmetry operations mixes different 
coordinates, then the first approach is not applicable. It is also not
applicable at all, if the primitive basis is not a direct product basis.

Since the symmetric top basis is a coupled basis, 
the first method mentioned above is not applicable. Therefore, 
the symmetry adapted DVR functions should be constructed by symmetry 
adapting the primitive DVR functions.

Before discussing the symmetry adaptation of the grid basis, let's define
$\vert \alpha_i \rangle $, $\vert \beta_j \rangle$, $\vert \gamma_k \rangle$,
  as the basis functions which are localized around the points $\alpha_i$,
$\beta_j$ and $\gamma_k$  for the angles $\alpha$, $\beta$, and $\gamma$ respectively.
Then,  
\begin{equation}
\vert \alpha_i \beta_j \gamma_k \rangle = \vert \alpha_i \rangle \vert \beta_j
 \rangle  \vert \gamma_k \rangle,
\end{equation}
 represents a direct product basis function in the grid basis. 
The transformation properties of the Euler angles given in table 
\ref{tab:monetr}
requires that if $\alpha_i$ and $\gamma_k$ are grid points than $\pi-\alpha_i=\bar{\alpha_i}$,
$\pi+\gamma_k=\gamma_k^*$, $-\gamma_k=\bar{\gamma_k}$ and $\pi-\gamma_k=\bar{\gamma_k}^*$, should also
be grid points. 

Since the group $C_{2v}(M)$ is the direct product of the group $G_2$
whose character table is given in table \ref{tab:chartabg2}
and the inversion group $\varepsilon$ whose character table is given in 
table \ref{tab:chartabeps}; symmetry adaptation of the 
basis functions can be done in two steps by first symmetry adapting
to the irreducible representations of the 
group $G_2$ and then symmetry adapting to the irreducible representations
of the group $\varepsilon$.

Effect of the $(12)$ operation to the Euler angles is given in  table
 \ref{tab:monetr}.
 With this transformation DVR basis functions symmetry adapted
to the $G_2$ permutation group  can be written as
\begin{equation}
  \vert \alpha_i \beta_j \gamma_k;s\rangle = \frac{1}{\sqrt{2}}
\left(    \vert \alpha_i \beta_j \gamma_k \rangle 
              + (-1)^s \vert \alpha_i \beta_j \gamma_k^* \rangle \right).
\end{equation}
In this equation, the parameter $s$ depends on the specific symmetry such that
$(-1)^s$ becomes the character of the $(12)$ operation in the irreducible 
representation that the basis function belongs to . 
The symmetry adaptation to the $G_2$ 
permutation group reduces the range of the angle $\gamma$
by a half so that the range of angle $\gamma$ becomes 
\begin{equation}
 0 \leq \gamma_k < \pi .
\end{equation}

Basis Functions can also be symmetry adapted to the inversion subgroup of 
the molecular symmetry group of the water dimer. In this case,
symmetry adapted basis functions will be in the form of 
\begin{equation}
\vert \alpha_i \beta_j \gamma_k;sl \rangle = N_l
 \left( \vert \alpha_i \beta_j \gamma_k;s \rangle
         + (-1)^l \vert \bar{\alpha_i} \beta_j \bar{\gamma_k};s \rangle \right),
\end{equation}
where the normalization constant is given by, 
\begin{equation}
N_l =\frac{1}{\sqrt{2(1+\delta_{\alpha_i \bar{\alpha_i}}\delta_{\gamma_k \bar{\gamma_k}})}}.
\end{equation}

The symmetry adaptation to the inversion subgroup of the molecular symmetry
group of the dimer reduces the range of $\alpha$ by
a half so that the range of $\alpha$ becomes 
\begin{equation}
\frac{\pi}{2}\leq \alpha_i \leq \frac{3\pi}{2}.
\end{equation}

The symmetry adapted DVR functions can be obtained for each irreducible
representation of the group $C_{2v}(M)$ by taking different combinations
of the parameters $s$ and $l$. 
The values of the parameters $s$ and $l$ are given in table 
\ref{tab:valsp} for
each irreducible representation.

\begin{table}
\caption[Partitioning of the symmetry adapted grid basis of a monomer
for water dimer calculations.]
{Values of the parameters $s$ and $l$ for the symmetry adapted
grid basis.}
\label{tab:valsp}
\begin{center}
\begin{tabular}{c|cc}
Representation & $s$ & $l$ \\ \hline
$A_1$ & $0$ & $0$ \\
$A_2$ & $0$ & $1$ \\
$B_1$ & $1$ & $1$ \\
$B_2$ & $1$ & $0$ 
\end{tabular}
\end{center}
\end{table}

Before closing this section a few points about the partitioning of the 
grid basis should be mentioned. The grid points in $\beta$ are obtained as the DVR points 
of Gauss-Legendre 
DVR of $\cos \beta$ 
by using the standard procedure \cite{jcl160,jcl78}. The
DVR points in $\cos \beta$ are distributed symmetrically 
with respect to $\beta=\pi/2$.

In the case of the angles $\alpha$ and $\gamma$, grid points are evenly and  
periodically distributed between $0$ and $2\pi$. Thus, the grid points of 
the angles $\alpha$ and $\gamma$ which are the plane wave DVR points
 are given by 
\begin{equation}
   \phi_j=j\left( \frac{2\pi}{N} \right) 
   \label{phigridpoints}
\end{equation} 
where $\phi$ is either $\alpha$ or $\gamma$,  $N$ is the number of DVR points and $j=0,1,\ldots,N-1$. According to this
prescription $0$ is always a grid point. As a result of that, the 
transformation properties of Euler angles given in table \ref{tab:transeu},
requires that $\pi$ should always be a grid point, too. 
According to equation (\ref{phigridpoints}), 
this is possible only if $N$ is an even number. Therefore, $N$ should 
be an even number in calculations.
 
\subsection{Transformation Matrix And The Evaluation of \\
            Potential Energy Matrix Elements}
\label{sec:tmatel}
The use of DVR provides a simple way for the evaluation of potential energy,
since the potential matrix is diagonal in the DVR basis. However, in the case
of coupled bases FBR basis is more compact, so that it is the primary basis
used in the calculations. Consequently, one needs to transform the potential
matrix from the DVR basis to the FBR basis, so that the total Hamiltonian
can be evaluated as 
\begin{equation}
 H^{(FBR)}=K^{(FBR)}+V^{(FBR)}=K^{(FBR)}+ T^{\dag} V^{(DVR)} T .
\end{equation}

The way to define the transformation matrix between the symmetric top basis 
and the corresponding grid basis is discussed by Leforestier 
\cite{leforest}. The transformation matrix elements between the symmetric
top basis, $\symt$, and the grid basis are given by 
\begin{equation}
T^{jkm}_{\alpha_i \beta_j \gamma_k}= \sqrt{\frac{2j+1}{2}} 
				\frac{e^{i m \alpha_i}}{\sqrt{N_{\alpha}}}
               \frac{e^{i k \gamma_k}}{\sqrt{N_{\gamma}}}
				\sqrt{w_{\beta_j}}d^j_{mk}(\cos \beta_j),
\label{tmat}
\end{equation}
where $N_\alpha$ and $N_\gamma$ are the number of grid points for 
the angles $\alpha$ and $\gamma$, $w_{\beta_j}$ is the weight 
of the grid point $\beta_j$ of the Gauss-Legendre quadrature for 
$\cos \beta$. 
Thus, the DVR basis functions are defined as 
\begin{eqnarray}
\delta_{\alpha_i \beta_j \gamma_k}(\alpha, \beta, \gamma) & = & 
                       \sum_{jkm} \symt T^{jkm}_{\alpha_i \beta_j \gamma_k} \\
& = & \sum_{jkm}
\frac{1}{2\pi}\sqrt{\frac{2j+1}{2}}\wigs T^{jkm}_{\alpha_i\beta_j\gamma_k} \\
&=& \sum_{jkm}\frac{2j+1}{4\pi} 
				\frac{e^{i m \alpha_i}}{\sqrt{N_{\alpha}}}
               \frac{e^{i k \gamma_k}}{\sqrt{N_{\gamma}}}
				\sqrt{w_{\beta_j}}d^j_{mk}(\cos \beta_j) \wigs .
	\nonumber \\
\end{eqnarray}

The transformation matrix can be decomposed into three one dimensional 
transformations:
\begin{equation}
T^{jkm}_{\alpha_i \beta_j \gamma_k}= K^{jkm}_{\beta_j} L_{\alpha_i}^m M_{\gamma_k}^k, 
\end{equation}
 where one dimensional transformations are defined as
\begin{eqnarray}
K^{jkm}_{\beta_j} & = &\sqrt{\frac{2j+1}{2}} \sqrt{w_{\beta_j}} d^j_{mk} (\cos \beta_j), \\
 L_{\alpha_i}^m & = & \frac{e^{i m \alpha_i}}{\sqrt{N_{\alpha}}}, \\
M_{\gamma_k}^k & = &  \frac{e^{i k \gamma_k}}{\sqrt{N_{\gamma}}} .
\end{eqnarray}

On the other hand, the transformation matrix is not a direct product of the one 
dimensional transformations given above, because the transformation matrix
describing the transformation form the basis of $d^j_{mk}(\cos \beta)$
 functions to the grid basis in  
$\cos \beta$ depends parametrically to the 
values of $k$ and $m$. For this reason, the transformation from the 
$d^j_{mk}(\cos \beta)$ functions to the grid in $\cos \beta$ should be done first. Then, the
resulting intermediate basis becomes decoupled, and the transformation from
this intermediate basis to the three dimensional grid basis becomes a two 
dimensional Fourier transformation.
This idea is suggested and used by Leforestier and co-workers 
in their water dimer paper \cite{leforestier1}.

\section{Generation of Bases for The Second Monomer}

Since the bases of the second monomer will be generated from the bases of 
the first monomer. The primitive functional 
basis of the second monomer will also be 
Wigner rotation functions. Thus, the primitive basis of the angular problem
will be the direct product of two symmetric top bases: $\symt \otimes \symt$. 
These basis functions can be written in terms of Wigner rotation functions as
\begin{equation}
\pbas =\frac{\sqrt{(2j_1+1)(2j_2+1)}}{8 \pi^2}  
	D^{j_1*}_{m_1k_1}(\alpha_1,\beta_1,\gamma_1) 
	D^{j_2*}_{m_2k_2}(\alpha_2,\beta_2,\gamma_2) 
\end{equation}
where $(\alpha_i,\beta_i,\gamma_i)$ are the Euler angles of the monomer i 
with respect to the body fixed frame of the dimer.

In order to find the effect of the symmetry operations to the basis 
functions, it is necessary to find
 the transformation properties of Euler angles under 
the action of symmetry operations. The effect of the symmetry operations to the 
Euler angles are given by Althorpe and Clary \cite{clary2}. However,
their definition of the body fixed axis for the monomers is different from
the one used here. In their definition the water molecule is in the $zy$ plane,
while here the water molecule is in the $xz$ plane. For that
reason, the transformation properties of 
the Euler angles are re-derived. The difference
between the transformations they have and the transformations given here is
only in the effect of the $E^*$ operation. Transformation properties 
of Euler angles is given in table \ref{tab:transeu}. For a discussion
of the derivations see appendix \ref{app:transeu}.

\begin{table}
\caption{Transformation Properties of Euler angles under the
effect of the symmetry 
operations of the $G_{16}$ permutation inversion group.}
\label{tab:transeu}
\begin{center}
\begin{tabular}{cll}
\hline \hline 
$E$ & $\alpha_1,\beta_1,\gamma_1$ & $\alpha_2,\beta_2,\gamma_2$ \\
\hline 
$E^*$ & $\pi-\alpha_1, \beta_1, -\gamma_1$ & $\pi-\alpha_2,\beta_2,-\gamma_2$ \\
$(12)$ & $\alpha_1,\beta_1,\pi+\gamma_1$ & $\alpha_2,\beta_2,\gamma_2$ \\
$(34)$ & $\alpha_1,\beta_1,\gamma_1$ & $\alpha_2,\beta_2,\pi+\gamma_2$ \\
$(ab)(13)(24)$ & $-\alpha_2,\pi-\beta_2,\pi+\gamma_2$ & $-\alpha_1,\pi-\beta_1,\pi+\gamma_1$ \\ 
\hline \hline
\end{tabular}
\end{center}
\end{table}

By using the symmetry properties of the Wigner rotation functions,
 the effect of the 
symmetry operations to primitive basis functions can be found easily.
The results are summarized in table \ref{tab:transsy}.

\begin{table}
\caption{Effects of the symmetry operations to symmetric top basis 
functions.}
\label{tab:transsy}
\begin{center}
\begin{tabular}{ll}
\hline \hline
$E$ & $\pbas $ \\
\hline
$E^*$ & $ (-)^{k_1+k_2}\vert j_1 \bar{k_1} \bar{m_1}j_2 \bar{k_2}\bar{m_2} \rangle $ \\
$(12)$ & $(-)^{k_1} \pbas $ \\
$(34)$ & $(-)^{k_2} \pbas $ \\
$(ab)(13)(24)$ & $(-)^{j_1+j_2}\vert j_2k_2\bar{m_2}j_1k_1\bar{m_1}\rangle $ \\
\hline \hline
\end{tabular}
\end{center}
\end{table}

As discussed in  the first paper, the bases of the second 
monomer should be generated from the bases of the first monomer by 
using the generator of the group that contains the permutations of the
identical monomers. For water dimer the generator of the group that
contains the permutations of the monomers is the operation $(ab)(13)(24)$.
The effect of this operation to the Euler angles and to the primitive 
bases are given in tables \ref{tab:transeu} and \ref{tab:transsy},
respectively. 

If the $i^{\mathrm{th}}$
 basis function belonging to the irreducible representation $\Gamma$
of  the group $C_{2v}(M)$ of the monomer $a$ has the form

\begin{equation}
\Gamma_i^{(a)}= \sum_l C_i^l \vert j_lk_lm_l \rangle;
\end{equation} 
then, the corresponding basis function belonging to the representation $\Gamma$
of the group $C_{2v}(M)$ of monomer $b$ will have the form
\begin{equation}
\Gamma_i^{(b)}=(ab)(13)(24)\Gamma_i^{(a)}=\sum_l (-1)^{j_l} C_i^l 
\vert j_lk_l\bar{m_l} \rangle.
\end{equation}

Although, the equations above are written for primitive basis functions,
it is obvious that they apply to the symmetry adapted basis functions 
given in table \ref{parmonbas}.

It is also necessary to generate a grid basis for the monomer $b$, from the
grid basis of monomer $a$. This will be done in the same way that the
spectral basis of monomer $b$ is generated from the spectral basis of 
monomer $a$. Thus, in order to generate a grid basis for the monomer $b$
the operation $(ab)(13)(24)$ should be applied to the grid basis functions 
of the monomer $a$. 

If $\vert \alpha_i^{(a)},\beta_j^{(a)},\gamma_k^{(a)}\rangle$ 
is a grid basis function of monomer $a$, then the corresponding grid basis 
function, $\vert \alpha_i^{(b)},\beta_j^{(b)},\gamma_k^{(b)}\rangle$, of the  monomer $b$
will be
\begin{equation}
\vert \alpha_i^{(b)},\beta_j^{(b)},\gamma_k^{(b)}\rangle = (ab)(13)(24)
\vert \alpha_i^{(a)},\beta_j^{(a)},\gamma_k^{(a)}\rangle .
\end{equation}  

The transformation properties of the Euler angles, 
given in table \ref{tab:transeu}, says that the 
transformation properties of Euler angles under the operation of the 
permutation operation 
$(ab)(13)(24)$ is given by
\begin{equation}
(ab)(13)(24)(\alpha_1, \beta_1, \gamma_1, \alpha_2, \beta_2, \gamma_2)
= (-\alpha_2, \pi-\beta_2, \pi+\gamma_2, -\alpha_1, \pi-\beta_1, \pi+\gamma_1).
\end{equation}  
Therefore, if  a grid basis function of the monomer $a$ 
is localized around the point
$(\alpha_i, \beta_i, \gamma_i)$, then the corresponding grid basis function for
the monomer $b$ will be localized  around the point 
$-\alpha_i, \pi-\beta_i, \pi+\gamma_i$.  The effect of the permutation 
operation $(ab)(13)(24)$ to the Euler angles of the monomers is to relabel the 
 angles so that they belong to monomer 
$b$, and to change the point the basis function is localized. 
Thus, this operation also mixes the order of the DVR functions.

\section{Combining Monomer Bases}

\begin{table}
\caption{\label{tab:bastabg16}This table shows which monomer bases should
be combined for obtaining bases for the water dimer calculations with
the group $G_{16}$. In the table, labels of the irreducible representations
are used to imply basis functions belonging to that symmetry. For an explanation of how to obtain mutually orthogonal basis for the doubly degenerate levels
(i.e. $E^+_x$, $E^+_y$) see reference \cite{MEOPhd}.}

\begin{center}
\begin{tabular}{cccc}
\hline \hline
$G_{16}$ & Bases & $G_{16}$ & Bases \\ \hline
$A_1^+$  & $(A_1 \otimes A_1) \oplus (A_2 \times A_2)$ &
             $A_1^-$ & $(A_1 \otimes A_2) \oplus (A_2 \otimes A_1)$ \\
$A_2^+$  & $(B_1 \otimes B_1) \oplus (B_2 \otimes B_2)$ &
             $A_2^-$ & $(B_1 \otimes B_2) \oplus (B_2 \otimes B_1)$ \\
$B_1^+$  & $(A_1 \otimes A_1) \oplus (A_2 \otimes A_2)$ &
             $B_1^-$ & $(A_1 \otimes A_2) \oplus (A_2 \otimes A_1)$ \\
$B_2^+$  & $(B_1 \otimes B_1) \oplus (B_2 \otimes B_2)$ &
             $B_2^-$ & $(B_1 \otimes B_2) \oplus (B_2 \otimes B_1)$ \\
$E^+_x$ & $(A_1 \otimes B_2) \oplus (A_2 \otimes B_1)$ &
             $E^-_x$ & $(A_1 \otimes B_1) \oplus (A_2 \otimes B_2)$ \\
$E^+_y$ & $(B_2 \otimes A_1) \oplus (B_1 \otimes A_2)$ &
             $E^-_y$ & $(B_1 \otimes A_1) \oplus (B_2 \otimes A_2)$ \\
\hline \hline
\end{tabular}
\end{center}
\end{table}

In the first paper, the way to combine monomer bases for water
dimer is discussed and the bases that should be used for each symmetry
are found. The results are given in table \ref{tab:bastabg16}. 
Thus, after choosing the 
right bases for each symmetry, the basis functions can be symmetry 
adapted to an irreducible representation $\Gamma$ of the group 
$G_{16}$ which is the molecular symmetry group of the water
 dimer by application of 
the projection operator 
\begin{equation}
   P = \frac{1}{2} (E + \chi^{\Gamma}[(ab)(13)(24)]^* (ab)(13)(24)),
\label{eq:projopcmb}
\end{equation}
where $\chi^{\Gamma}[(ab)(13)(24)]$ is the character of the permutation
operation $(ab)(13)(24)$ in the irreducible representation $\Gamma$.
 These characters can be found in the character
table of the group $G_{16}$ which is given in table \ref{tab:chartabg16}.

\begin{table*}
\caption{\label{tab:chartabg16}Character table of the $G_{16}$ PI group. This group is isomorphic
to the $D_{4h}$ point group. In the table, $\Gamma = x \otimes y$ means that
the irreducible representation $\Gamma$ of the group $G_{16}$ is the direct
product of the irreducible representation $x$ of the group $G_8$, which is the
pure permutation subgroup of the water dimer,
 and the irreducible  
representation $y$ of the inversion 
group $\varepsilon$, whose charter table is given
in table \ref{tab:chartabeps}. This character table is taken from the
reference [\onlinecite{dyke77}]. The correlations between the irreducible
representations of the group $G_{16}$ with the irreducible representations
of its subgroups $G_8$ and $\varepsilon$ are added by the author.}

\begin{center}
\begin{tabular}{crrrrrrrrrr}
\hline \hline
 & & $(12)$ & $(ab)(13)(24)$ & $(ab)(1324)$ & & & $(12)^*$ & $(ab)(13)(24)^*$ & $(ab)(1324)^*$ & \\
$G_{16}= G_8 \otimes \varepsilon $ & $E$ & $(34)$ & $(ab)(14)(23)$& $(ab)(1423)$& $(12)(34)$ & $E^*$ &
            $(34)^*$ & $(ab)(14)(23)^*$ & $(ab)(1423)^*$ & $(12)(34)^*$ \\
\hline
$A_1^+ = A_1 \otimes G $ & $1$ & $1$ & $1$ & $1$ & $1$ & $1$ & $1$ & $1$ & $1$ & $1$  \\
$A_2^+ = A_2 \otimes G $ & $1$ & $-1$ & $-1$ & $1$ & $1$ & $1$ & $-1$ & $-1$ & $1$ & $1$  \\
$B_1^+ = B_1 \otimes G $ & $1$ & $1$ & $-1$ & $-1$ & $1$ & $1$ & $1$ & $-1$ & $-1$ & $1$  \\
$B_2^+ = B_2 \otimes G $ & $1$ & $-1$ & $1$ & $-1$ & $1$ & $1$ & $-1$ & $1$ & $-1$ & $1$  \\
$E^+ = E \otimes G$   & $2$ & $0$ & $0$ & $0$ & $-2$ & $2$ & $0$ & $0$ & $0$ & $-2$  \\
$A_1^- = A_1 \otimes U $ & $1$ & $1$ & $1$ & $1$ & $1$ & $-1$ & $-1$ & $-1$ & $-1$ & $-1$  \\
$A_2^- = A_2 \otimes U $ & $1$ & $-1$ & $-1$ & $1$ & $1$ & $-1$ & $1$ & $1$ & $-1$ & $-1$  \\
$B_1^- = B_1 \otimes U $ & $1$ & $1$ & $-1$ & $-1$ & $1$ & $-1$ & $-1$ & $1$ & $1$ & $-1$  \\
$B_2^- = B_2 \otimes U $ & $1$ & $-1$ & $1$ & $-1$ & $1$ & $-1$ & $1$ & $-1$ & $1$ & $-1$  \\
$E^- = E \otimes U $   & $2$ & $0$ & $0$ & $0$ & $-2$ & $-2$ & $0$ & $0$ & $0$ & $2$  \\
\hline \hline
\end{tabular}
\end{center}
\end{table*}

In order to illustrate symmetry adaptation of the basis functions, 
consider the irreducible representation $A_1^+$. Table \ref{tab:bastabg16}
says that the bases that should be used in the calculations should be 
$(A_1 \otimes A_1) \oplus (A_2 \otimes A_2)$.  Therefore, the basis that will be used in the calculations
of the $A_1^+$ levels is formed by taking a direct product of the $A_1$ basis
for monomer $a$ and the $A_1$ basis for monomer $b$ and combining this with the 
direct  product of the $A_2$ basis for monomer $a$  and the 
$A_2$ basis for monomer
$b$. If $A_{1i}^{(a)}$ is the $i^{\mathrm{th}}$ basis function 
which has the $A_1$ symmetry for the monomer $a$, and $A_{1i}^{(b)}$, 
$A_{2i}^{(a)}$,  $A_{2i}^{(b)}$ are defined similarly; then, the basis
functions of  $A_1^+$ calculation before the 
symmetry adaptation to the permutation group $G_2^{(ab)}$ will be in the form
of either
\begin{equation}
   \psi_{ij}^{(A_1^+)}= A_{1i}^{(a)} A_{1j}^{(b)},
\label{a1pfp}
\end{equation} 
or
\begin{equation}
   \psi_{ij}^{(A_1^+)}= A_{2i}^{(a)} A_{2j}^{(b)}.
\label{a1psp}
\end{equation} 
After the symmetry adaptation to the permutation group 
$G_2^{(ab)}$, the form of 
the basis functions will be either 
\begin{equation}
\Psi_{ij}^{(A_1^+)}= A_{1i}^{(a)} A_{1j}^{(b)} + A_{1j}^{(a)} A_{1i}^{(b)},
\end{equation}
or 
\begin{equation}
\Psi_{ij}^{(A_1^+)}= A_{2i}^{(a)} A_{2j}^{(b)} + A_{2j}^{(a)} A_{2i}^{(b)}.
\end{equation}
Thus, the symmetry adaptation reduces the sizes of both the 
$(A_1^{(a)} \otimes A_1^{(b)})$ basis
and the $(A_2^{(a)} \otimes A_2^{(b)})$
 basis to the half of their original size. The sign 
between the two terms on the right hand side of the equation is plus
because the character of the $(ab)(13)(24)$ operation in the $A_1^+$
representation is $+1$. 

Equations (\ref{a1pfp}) and (\ref{a1psp}) apply to the $B_1^+$ 
irreducible representation as well
as it can be seen from table \ref{tab:bastabg16}. However, symmetry adapted 
combinations will be different of course. Since the character of the
$(ab)(13)(24)$ operation for the $B_1^+$ operation is $-1$, symmetry adapted
basis functions will be in the form of either 
\begin{equation}
\Psi_{ij}^{(B_1^+)}= A_{1i}^{(a)} A_{1j}^{(b)} - A_{1j}^{(a)}A_{1i}^{(b)},
\end{equation}
or 
\begin{equation}
\Psi_{ij}^{(B_1^+)}= A_{2i}^{(a)} A_{2j}^{(b)} - A_{2j}^{(a)}A_{2i}^{(b)}.
\end{equation}

When the same procedure is applied to the $A_2^-$ representation, the
form of the basis functions before the symmetry adaptation becomes either
\begin{equation}
\psi_{ij}^{(A_2^-)} = B_{1i}^{(a)} B_{2j}^{(b)},
\end{equation}
or
\begin{equation}
\psi_{ij}^{(A_2^-)} = B_{2i}^{(a)} B_{2j}^{(b)}.
\end{equation}
After the symmetry adaptation one gets
\begin{equation}
\Psi_{ij}^{(A_2^-)}= B_{1i}^{(a)} B_{2j}^{(b)} - B_{2j}^{(a)}B_{1i}^{(b)}.
\end{equation}
Thus, in this case
  symmetry adaptation will mix the two different parts. So that 
one can take either the first product basis $(B_1^{(a)}\otimes B_2^{(b)})$
 or the 
second product basis $(B_2^{(a)} \otimes B_1^{(a)})$ and  symmetry adapt
it to get the full basis. Since symmetry adaptation mixes these two 
different product bases with each other, only one of them is sufficient to make
a symmetry adapted calculation.

Symmetry adapted bases for all of the other symmetries can be constructed
similarly. It should also be noted that since the character of the 
operation $(ab)(13)(24)$ is $0$ for the doubly degenerate levels, the 
projection operator given in equation (\ref{eq:projopcmb}) is nothing but
the identity operation. Therefore, once the basis functions are formed,
they are already symmetry adapted as they are 
and there is no further symmetry adaptation.
In table \ref{tab:bastabg16}, two separate bases are shown for the 
doubly degenerate levels which are labeled with subscripts $x$ and $y$. 
These bases are orthogonal to each other and they do not mix with each other.
In order to solve the eigenvalue problem for the doubly degenerate levels
either bases can be used. They will have the same spectrum. The fact that
there are two different bases for a doubly degenerate level that do not
mix with each other but still give the same set of eigenvalues is the
physical explanation of the double degeneracy, of course. 

\section{Angular Calculations}
\label{ssec:fdcal}
In order to solve the eigenvalue problem for five dimensional angular 
problem. It is necessary to evaluate the matrix elements of the angular
Hamiltonian given in equation (\ref{partition}). Since the contracted
basis functions of the monomers are already the eigenstates of the 
model Hamiltonians, their evaluation is easy. 
The term $\hat{\Delta K}$ in that equation can be evaluated in the primitive 
basis of the monomers easily, and then can be transformed to the 
contracted bases of the monomers by using the transformation matrix 
which is obtained by solving the eigenstates of the model Hamiltonian's
of the monomers. The term $\hat{\Delta V}$, can be evaluated easily
in the grid basis which is a tensor product of the monomer grid bases, and
then can be transformed to the contracted bases of monomers in two steps
first by transforming from grid basis to the primitive functional basis
and 
then transforming from the primitive functional basis 
to the angular basis which is the tensor product of the contracted bases of 
the monomers.  

\section{Radial Calculation}
\label{sec:radcal}
Once the angular problem is solved at several fixed $R$ values, the
eigenvalues for the full problem can be found by fitting the results of 
the angular calculations to a Morse function and
solving for the eigenvalues.
The Hamiltonian for this one dimensional problem becomes 
\begin{equation}
\hat{H}= -\frac{1}{2\mu R} \frac{\partial^2}{\partial R^2} + V(R),
\label{radialhamil}
\end{equation}
where $V(R)$ is the Morse function which approximates the eigenvalues of 
the angular calculations at the given $R$ values.
Since the Morse potential, given in the form, 
\begin{equation}
V(r)= D(e^{-2\alpha(r-r_0)}-2 e^{-\alpha (r-r_0)}),
\end{equation}
includes three parameters, $D$, $\alpha$, $r_0$; it is sufficient to solve
 the angular problem at three different $R$ values. If only three points are
used, there exist a unique function which fits to the given data. If the 
calculation is done at more than three points, then the Morse function which
fits to the data can be found by making a least squares fit.

When the Morse fit to the function is done, the eigenvalues of the
Hamiltonian, 
given in equation (\ref{radialhamil}), can be found easily; since 
for $J=0$, the eigenvalues are analytic. According to Landau and Lifschitz,
the eigenvalue of the $n^{\mathrm{th}}$ level  is given by \cite{Landau3}
\begin{equation}
E_n=-D\left(1- \frac{\alpha}{\sqrt{2 \mu D}}(n+\frac{1}{2}) \right)^2,
\end{equation}
where $n$ takes integral values from zero to the greatest value
for which the expression in the parentheses is positive.

\section{Details of Calculations}

The calculations are done by using the SAPT-5st 
potential surface 
developed by Groenenboom {\it et. al.} \cite{avoirdpes00,avoirdpes00a}. 
The source code of this potential surface was made available to 
public by Groenenboom {\it et. al.} \cite{avoird-prl}
as an EPAPS document with the document number EPAPS:E-PRLTAO-84-060018.
The source code can be obtained via ftp from the site ftp.aip.org under 
the directory /epaps/ . The mass  of {\WA} is taken as $18.010560$
and the moments of inertia of water monomers 
are taken as 
$A=27.8806 \mathrm{cm}^{-1}$, $B=14.5216 \mathrm{cm}^{-1}$ and 
$C=9.2778 \mathrm{cm}^{-1}$. 
 These values are the same with the values 
that are used in the original calculations of Groenenboom { \it et. al.} 

While doing the calculations primitive functional basis of the monomers 
are taken as symmetric top basis 
with $j \leq 10$ and $m \leq 8$. Before 
the symmetry adaptation this corresponds to a basis size of $\approx 1650$.
The number of the grid points in $\alpha$ and $\gamma$ are set to $26$
and the number of grid points in beta are set to $15$ before symmetry 
adaptation. All of the calculations are done for $J=0$.  

In monomer calculations, the spectral basis is fully symmetry adapted and
the grid basis is symmetry adapted to the permutation of the protons but
not to the inversion symmetry. The matrix representing the Hamiltonian 
operator in the symmetry adapted symmetric top basis is stored in memory, and
the diagonalization is done directly.  

While doing the angular calculations, the angular basis which is obtained
as a tensor product of the contracted bases of the monomers is not 
symmetry adapted to the full symmetry of the water dimer. Instead, 
the calculations are done by using the Symmetry Adapted Lanczos (SAL)
algorithm \cite{Carr-sal}. 
 The use of the SAL algorithm
allows 
one to diagonalize more
than one symmetries at once. In the case of water dimer problem being 
considered here, SAL method made it possible to solve for the eigenvalues
of the $A_1^+$ and $B_1^+$ levels together, and also $A_2^-$ and $B_2^-$
levels together. This results from the fact that the angular basis 
of the $A_1^+$ and $B_1^+$ levels and similarly the angular basis of the
$A_2^-$ and $B_2^-$ levels are the same before symmetry adaptation 
(see table \ref{tab:bastabg16}). In the case of doubly degenerate levels
calculations should be done separately for each level since the bases of 
double degenerate levels are unique to themselves. However, use of the
SAL algorithm still makes the calculations faster since in the SAL algorithm
projection operators are used to get the symmetry adapted eigenfunctions. 

Angular calculations are done at three different fixed $R$ values 
which are $5.38 \mathrm{a.u.}$, $5.53 \mathrm{a.u.}$ and $5.68 \mathrm{a.u.}$
The ground states eigenvalues are used to define a potential surface for 
the stretching motion. The potential surface of the stretching motion is
found by making a nonlinear fit to the Morse function by using Newton's
algorithm \cite{isaacson}. 

In order to converge the results it was necessary
to use $100$ basis functions per monomer for the angular calculations.

\section{\label{ssec:res}Results And Discussions}
\begin{table}
\caption[A comparison of the results of MBR calculations with the results of 
Groenenboom {\it et. al.}]
{\label{tab:results} A comparison of the results of MBR calculations
with the results
of Groenenboom { \it et. al.} The results of Groenenboom {\it et. al.}
 are taken from 
the table III of the reference \cite{avoirdpes00a}. The results in the
table are in $\mathrm{cm^{-1}}$. MBR results are obtained by using $100$
basis functions per monomer. In the table $a$ is the
splitting due to acceptor tunneling; $i_1$ and $i_2$ are the splittings 
between $A_1^+ /B_1^+$ and $A_2^-/B_2^-$ levels due to interchange tunneling.} 
\begin{center}
\begin{tabular}{ccc}
\hline \hline 
Symmetry & Groenenboom { \it et. al.} & MBR \\ \hline
$A_1^+$  & $-1076.8643$ &  $-1075.2116$ \\
$E+$     & $-1076.4312$ &  $-1074.8698$ \\
$B_1^+$  & $-1076.1419$ &  $-1074.4688$ \\
$A_2^-$  & $-1065.6333$ &  $-1063.0106$ \\
$E^-$    & $-1065.2540$ &  $-1062.6818$ \\
$B_2^-$  & $-1064.9825$ &  $-1062.3926$ \\ 
$a$      & $11.19$      &  $12.14$      \\
$i_1$    & $0.722$      &  $0.743$      \\
$i_2$    & $0.651$      &  $0.618$      \\
\hline \hline
\end{tabular}
\end{center}
\end{table}

\begin{figure*}
\caption{\label{fig:conv} Convergence of the results of the MBR 
calculations with the number of  angular
basis functions per monomer.} 
\begin{center}
\input{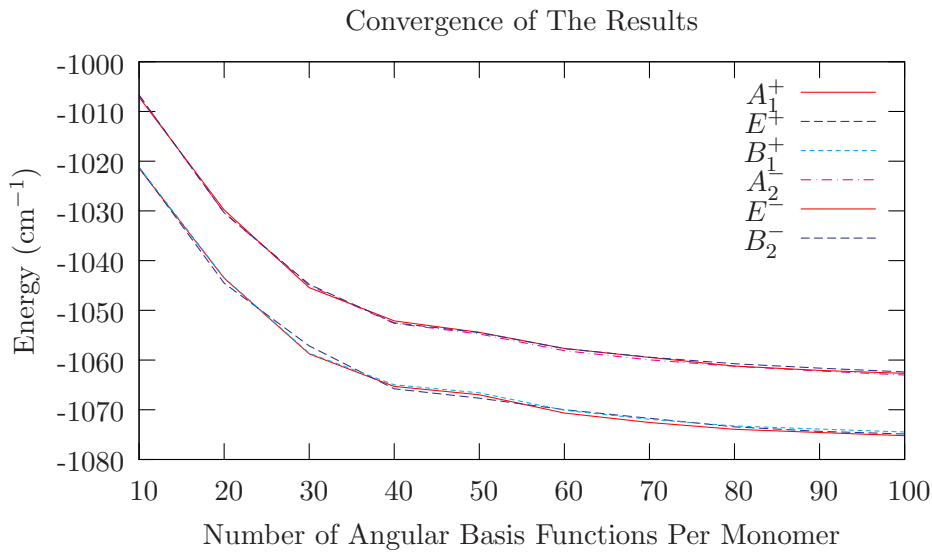}
\end{center}
\end{figure*}

\begin{figure}
\caption[A comparison of the results of the MBR calculations with the original
calculations of Groenenboom {\it et. al.} and the experimental data]
{\label{fig:results} A comparison of the results of MBR 
calculations (lower numbers)
with the original calculations of Groenenboom {\it et. al.} (middle numbers)
and the experimental data (upper numbers). Experimental data 
is not available for the acceptor switching.}
\begin{center}
\input{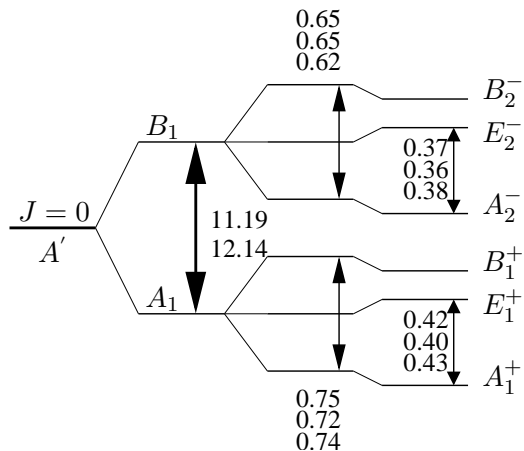}
\end{center}
\end{figure}

 A comparison of the MBR results with the original calculations of
Groenenboom {\it et. al.} is given in table \ref{tab:results}.  In the
table, $i_1$ is the tunneling splitting due to interchange tunneling 
between the $A_1^+$ and 
$B_1^+$ levels which is calculated as 
\begin{equation}
   i_1= E(B_1^+) - E(A_1^+),
\end{equation}
where $E(x)$ denotes the energy of the level $x$;
$i_2$ is the interchange tunneling between the $A_2^-$ and 
$B_2^-$ levels which is calculated as 
\begin{equation}
   i_2= E(B_2^-) - E(A_2^-);
\end{equation}
and $a$ is the tunneling splitting due to acceptor switching which is 
calculated as  
\begin{equation}
   a =  \frac{E(A_1^+) + E(B_1^+)}{2} - \frac{E(A_2^-) + E(B_2^-)}{2}.
\end{equation}
As it can be seen from the table the results are in good agreement with 
each other. Especially, the tunneling splittings are in very good 
agreement. From the table, it can be seen that the MBR calculation leads
to eigenvalues which are higher than the results of Groenenboom {\it et. al.}
This can be attributed to the fact that the stretching coordinate 
is treated in different ways in two calculations. In the calculations of 
Groenenboom {\it et. al.} the stretching coordinated is handled with a DVR 
grid with $49$ equally spaced points \cite{avoirdpes00a}.
 On the other hand, in the
MBR calculations stretching coordinate is separated from the angular 
coordinates adiabatically. Therefore, because of this difference
the calculations of Groenenboom {\it et. al.} are less approximate than the MBR 
calculations.  The adiabatic separation of the stretching coordinate from
the angular coordinates was first done by Althorpe and Clary \cite{clary2}. 
Their calculations are done with a different potential surface. 
A comparison of their results with other less approximate calculations 
which is done with the same potential surface is available \cite{jcl155}.
The comparisons shows that adiabatic approximation is successful in 
predicting the tunneling splittings. The MBR results also shows that
it is possible to get good results with adiabatic approximation. 

Convergence of the results with the number of basis functions per monomer
that is used in the angular calculations is shown in  figure
\ref{fig:conv}. The comparisons of the MBR results with the 
results of Groenenboom {\it et. al.} and the experimental data 
\cite{fraser91,zwart91,braly2} is also 
shown schematically in figure \ref{fig:results}.

From the comparison of the MBR results with the original results, it can 
be said that the MBR method gives good results. Since the number of 
the optimized basis functions that are used for each monomer $(100)$
is much more
smaller than the number of the primitive basis functions $(\approx 1650)$,
 it can also be
said that the method is efficient. 

If the calculations were done with the same primitive bases but without 
generation of any optimized bases, it would be possible to decrease the
size of the basis by a factor of $16$ with the help of standard symmetry
adaptation procedures since the molecular symmetry group of water dimer
is of the order of $16$. On the other hand, the use of the MBR method 
decreases the size of the basis of 
a single monomer with almost the same factor so that
the size of the cluster basis becomes about $16$ times smaller than 
what one would be able to achieve with standard symmetry adaptation 
procedures. 

The calculations given here can be improved in several ways. Firstly,
the stretching coordinate can be treated more accurately. This can be
achieved either by using more points to find the Morse potential or 
using more exact ways to handle it as Groenenboom {\it et. al.} have done
or by using sequential diagonalization truncation schemes. However,
this does not really seem to be necessary since the results are
quite successful. Secondly, from figure \ref{fig:conv}, it can be
seen that the convergence of the results is not uniform. The changes in
the results when the number of angular basis functions per monomer is 
increased from $50$ to $60$ are greater than the changes in 
the results when the number of angular basis functions per monomer 
is increased from $40$ to $50$. This shows that simply taking the states
with the lowest energies as a contracted basis is not a good idea. It 
might be possible to devise better strategies while forming the contracted
bases in order to obtain the best possible contracted basis. Although, 
this does not seem to be a big problem for a  six dimensional 
system, it might be important
when one wants to study higher dimensional systems.

\section{Conclusions}

Application of the MBR method has been illustrated by calculating the VRT 
spectra of water dimer by using the SAPT-5st potential surface of
Groenenboom {\it et. al.} \cite{avoirdpes00a}. The calculations are 
done by using Wigner rotation functions as primitive bases. The use 
of the MBR method made it possible to decrease size of monomer bases
by a factor of $\approx 16$. 
 The results of the calculations are in 
good agreement with both the original calculations of Groenenboom {\it et. al.}
and also with the experimental results. A detailed discussion of the results
can be found in section \ref{ssec:res}. 

 Because of its efficiency,
the MBR method can be used for studies of clusters bigger than dimers.
Thus, it can be used for studying the many-body terms and 
for deriving accurate potential surfaces. 
The results of calculations in this paper are especially 
encouraging for a study of water trimer with a pairwise 
potential surface. 
 For example: consider the nine dimensional
Hamiltonian derived by  van der Avoird {\it et. al.} \cite{trimerI:96}.
 In that model, all of the monomers are
allowed to rotate around their center of masses but
the center of masses of monomers are fixed in space. If the trimer calculation would require about the same
number of contracted basis functions for a monomer, then a study of the 
nine dimensional angular problem of water trimer would require $10^6$
basis functions.
Although, a problem of that size can be handled with iterative methods,
it is still quite big. Nevertheless, 
it is reasonable to expect that the trimer problem can be solved with 
less number of optimized basis functions per monomer. Firstly, 
water trimer is much more symmetric than water dimer. Secondly,
a study of water trimer with a pairwise potential surface
 will have much deeper potential 
wells. Because of the importance of the three-body
terms in the potential surface of water trimer, a calculation with 
a pairwise potential surface cannot give an accurate estimation of the 
experimental data. Nevertheless, this should be a big step towards the 
derivation of three-body terms in the potential surface of bulk liquid and
solid water.  
A qualitative model for a possible
application of the MBR method to water trimer is already given in the
first paper.
Its implementation remains as a future work. 

\appendix

\section{A Simple Way to Find The Transformation Properties of 
             Euler Angles}
\label{app:transeu}
The transformation properties of the Euler angles of  a body fixed frame under 
rotations is well know if the Euler angles are defined with respect to 
a space fixed frame, which does not move under the effect of 
any symmetry operation.
These transformations are given in the  book {\textit{Molecular Symmetry
and Spectroscopy}}, written by Bunker and Jensen \cite{bunker98}.
 For convenience of the reader,
they are given  in table \ref{eulertrans}.
However, the case of the small clusters is different, because the Euler 
angles of the body fixed frames of the
 monomers are defined with respect to 
the body fixed frame of the cluster, and not with respect to a space 
fixed frame. Since the body fixed frame of the cluster is not fixed 
in space, it also rotates with the effects of permutation inversion operations.
Therefore, the application of the transformations given in table
 \ref{eulertrans} is sufficient only if the body fixed frame of the cluster
does not move. However,  they will give wrong answers if  the body fixed frame 
of the cluster moves, too. For example, in the case of water dimer, 
transformations given in table \ref{eulertrans} will give the right answer for
the operation of $(12)$. However, they will not work for the operations 
$E^*$ and $(ab)(13)(24)$, because these operations rotate the body 
fixed frame of the dimer around its $x$ axis by $\pi$ radians. 

\begin{table}
\caption[Transformation properties of Euler angles under
the effect of the rotation of the body fixed frame of a monomer.]
{Transformation properties of Euler Angles. This table is 
taken from page 266 of the book {\it Molecular Symmetry and Spectroscopy}
written by Bunker and Jensen \cite{bunker98}. Coordinates are relabeled  
to make it consistent with notation used here. In the table,
 $R_{\phi}^{\pi}$ is a rotation of the molecule fixed $(x,y,z)$
axes through $\pi$ radians about an axis in the $xy$ plane making an 
angle $\phi$ with the $x$ axis ($\phi$ is measured in the right hand sense
about the $x$ axis), and $R_z^{\theta}$ is a rotation of molecule fixed 
$(x,y,z)$ axes through $\theta$ radians about $z$ axis ($\theta$ is measured
in the right hand sense about the $z$ axis). }
\label{eulertrans}
\begin{center}
\begin{tabular}{ccc}
\hline \hline
 & $R_{\phi}^{\pi}$ & $R_z^{\theta}$ \\
\hline 
$\beta$ &  $\pi -\beta$ & $\beta$ \\
$\alpha$ & $\alpha+\pi$ & $\alpha$ \\
$\gamma$ & $2\pi-2\phi-\gamma$ & $\gamma+\theta$ \\ 
\hline  \hline
\end{tabular}
\end{center}
\end{table}

To the best of author's knowledge there is not any easy way of finding the
transformations in these cases available in the literature.
 In the following lines, an easy way of finding these transformations
will be discussed. 

 In order to find the changes in the 
Euler angles, the process will be divided into two steps. In the first step
the orientation of the body fixed frame of the monomers will be kept fixed 
with respect to a space fixed frame. However, the body fixed frame 
of the cluster will be 
rotated together with the body, of course. In the second step, the rotation
of the monomer frames will be done.  

The transformations in the second step will be the same with the ones
given in table \ref{eulertrans}, because the cluster frame does not move in 
the second step. Therefore, if one finds the transformations for
the first step, then the overall transformation can be found by applying two 
transformations sequentially. 

The transformation properties of the first step can be found easily by realizing
that the only difference in the first step is that the roles of the 
two frames are interchanged. This time, it is the monomer frame which 
acts like the space fixed frame since it doesn't move; and it is the cluster 
frame that rotates. If $(\alpha, \beta, \gamma)$ are the Euler angles of the 
monomer frame defined with respect to the cluster 
frame, one can equally say that
$(-\gamma, -\beta, -\alpha)$ are the Euler angles of the cluster frame defined 
with respect to the monomer frame. Thus, the Euler angles of the cluster frame
with respect to the monomer frame after the rotation can be found 
by applying the transformations in table \ref{eulertrans}. Then,
 the Euler angles
of the monomer frame with respect to the 
cluster frame can be found by using the same
trick again. If $(\alpha',\beta',\gamma')$ are the Euler angles
of the cluster frame 
obtained by applying the transformations in table \ref{eulertrans}
to the Euler angles $(-\gamma,-\beta,-\alpha)$, then the Euler angles of the
monomer frame will be $(-\gamma', -\beta', -\alpha')$.

Thus, by using the transformation rules given in table \ref{eulertrans},
one can find the transformation properties of Euler angles of the monomer 
when it is only the frame of the cluster that rotates. The general
formulas for such transformations are derived 
and the results are summarized in table
\ref{passtrans}.  

\begin{table}
\caption[Transformation properties of Euler angles under the
effect of the rotation of the body fixed frame of the cluster.]{Transformation properties of Euler Angles. Definitions of
the rotation operations are the same with table \ref{eulertrans}. 
However, this time it is the cluster frame rotating, not the monomer frame.}
\label{passtrans}
\begin{center}
\begin{tabular}{ccc}
\hline \hline
 & $R_{\phi}^{\pi}$ & $R_z^{\theta}$ \\
\hline 
$\beta$ &  $\pi -\beta$ & $\beta$ \\
$\alpha$ & $2\phi-\alpha$ & $\alpha-\theta$ \\
$\gamma$ & $\pi+\gamma$ & $\gamma$ \\ 
\hline \hline
\end{tabular}
\end{center}
\end{table}

Since the order of the two successive operations 
that are used to find the transformation properties of the Euler angles
of the monomer 
is not important for the final orientation of the frames, the order in 
which one uses the transformations is not important.  

To illustrate the applications of the transformations given in tables 
\ref{eulertrans}, and \ref{passtrans}, to molecular clusters, consider
water dimer. Body fixed frames of the dimer and the monomers are given 
in figures \ref{fig:bffdimer} and \ref{monff}, respectively. 
Firstly, consider the effect of $E^*$ operation to  
water dimer. This operation rotates the body fixed frame of 
the monomer around its $y$ axis by $\pi$ radians (see figure \ref{monff})
 and  it also rotates
the body fixed frame of the dimer around its $x$ axis by $\pi$ radians
(see figure \ref{fig:bffdimer}).
By using the transformations in table \ref{eulertrans} (with $\phi=\pi/2$),
the  effect of the monomer
rotation to the Euler angles is found to be 
\begin{equation}
(\alpha, \beta, \gamma) \rightarrow (\pi+\alpha, \pi-\beta, \pi-\gamma). 
\end{equation}
Then, by using the transformations in table \ref{passtrans} (with $\phi=0$)
the effect of rotation of the dimer frame to the Euler angles of the monomer
frame  is found to be
\begin{equation}
(\pi+\alpha, \pi-\beta , \pi-\gamma) \rightarrow (\pi-\alpha, \beta, -\gamma,)
\end{equation}
where $-\alpha-\pi$ is replaced with $\pi-\alpha$. This is possible because
adding $2\pi$ to the angles $\alpha$ and $\gamma$ doesn't change anything 
since they have a period of $2\pi$.
Therefore, the overall transformation becomes,
\begin{equation}
(\alpha, \beta, \gamma) \rightarrow (\pi-\alpha, \beta, -\gamma).
\end{equation} 
This is what is reported in table \ref{tab:transeu}. 
The same transformation applies to both of the monomers. 

In the case of the 
operation $(ab)(13)(24)$, the effect of the operation to the monomer 
frames is just to relabel them, so the first transformation becomes

\begin{equation}
(\alpha_1,\beta_1,\gamma_1,\alpha_2, \beta_2, \gamma_2) \rightarrow 
(\alpha_2, \beta_2, \gamma_2, \alpha_1,\beta_1, \gamma_1).
\end{equation} 
This operation also rotates the dimer frame around its $x$ axis by $\pi$
radians, so with the transformations given in table \ref{passtrans}, 
one gets
\begin{equation}
(\alpha_2, \beta_2, \gamma_2, \alpha_1, \beta_1, \gamma_1) \rightarrow
(-\alpha_2, \pi-\beta_2, \pi+\gamma_2, -\alpha_1, \pi- \beta_1, \pi+ \gamma_1).
\end{equation} 
Therefore, the overall effect of the $(ab)(13)(24)$ operation to the 
Euler angles of the monomers become
\begin{equation}
(\alpha_1, \beta_1, \gamma_1, \alpha_2, \beta_2, \gamma_2) \rightarrow
(-\alpha_2, \pi-\beta_2, \pi+\gamma_2, -\alpha_1, \pi- \beta_1, \pi+ \gamma_1).
\end{equation} 

The symmetry operations $(12)$ and $(34)$ do
 not have any effect on the body fixed 
frame of the dimer. Therefore, their effect to the Euler angles of the 
monomers can be found by using the transformations given in table 
\ref{eulertrans}. The results will be the ones that are reported in table
\ref{tab:transeu}.

\section{Real Symmetric Top Basis}
\label{app:rstb}

The fact that the symmetric top basis is complex, makes the calculations
more demanding, since
 complex numbers occupies twice the memory real numbers
occupies. For this reason, it is advantageous to transform the symmetric 
top basis to a real basis that we will be called 
real
 symmetric top basis.
The way to generate a real basis from a complex basis is obvious: sum and 
differentiate with its complex conjugate. By using the symmetry properties of
the Wigner rotation functions \cite{rose}, it can be shown easily that
\begin{equation}
	\wigs=(-1)^{m-k} D^j_{-m-k}(\alpha, \beta, \gamma).
\end{equation}

From the equation above, 
it follows that $\symt^{*}= (-1)^{m-k}\symtb$. This leads
to the result that the 
linear combinations $\symt \pm \symtb$ are either pure real
or pure imaginary. In the case of pure imaginary functions, the imaginary number
$i$ can be omitted since it is just a phase factor.  After doing some 
algebra, following functions are obtained as the real symmetric top basis.

For $m-k$ is odd: 
\begin{eqnarray}
\frac{\symt + \symtb}{i\sqrt{2+2\delta_{k0}\delta_{m0}}} 
& = &  \frac{\sqrt{2j+1}}{2\sqrt{1+\delta_{k0}\delta_{m0}}}
			\sin(m\alpha+k\gamma) d^j_{mk}(\cos \beta), 
\label{real1} \\
\frac{\symt - \symtb}{\sqrt{2}} & = &
\frac{\sqrt{2j+1}}{2\sqrt{1+\delta_{k0}\delta_{m0}}}
                        \cos(m\alpha+k\gamma) d^j_{mk} (\cos \beta).
\label{real2}
\end{eqnarray}

For $m-k$ is even:
\begin{eqnarray}
\frac{\symt - \symtb}{i\sqrt{2}} 
& = &  \frac{\sqrt{2j+1}}{2\sqrt{1+\delta_{k0}\delta_{m0}}}
                        \sin(m\alpha+k\gamma) d^j_{mk}(\cos \beta),
\label{real3} \\
\frac{\symt + \symtb}{\sqrt{2+2\delta_{k0}\delta_{m0}}} & = &
\frac{\sqrt{2j+1}}{2\sqrt{1+\delta_{k0}\delta_{m0}}}
                        \cos(m\alpha+k\gamma) d^j_{mk}(\cos \beta).
\label{real4}
\end{eqnarray}

When the real symmetric top basis functions are used 
as a basis,
  the transformation matrix 
elements from this basis to the grid basis also becomes real. Thus,
by using equation (\ref{tmat}), transformation matrix elements for  
the basis functions given by equations (\ref{real1}) and (\ref{real3}) becomes 
\begin{equation}
T^{jkm}_{\alpha_i \beta_j \gamma_k}= 
	\frac{\sqrt{2j+1}}{2\sqrt{1+\delta_{k0} \delta_{m0}}}
                                \frac{\sin( m \alpha_i+ k \gamma_k)}
				{\sqrt{N_{\alpha} N_{\gamma}} }
                                \sqrt{w_{\beta_j}}d^j_{mk}(\cos \beta_j),
\end{equation}
and the transformation matrix elements for the basis functions given in 
equations (\ref{real2}) and (\ref{real4}) becomes
\begin{equation}
T^{jkm}_{\alpha_i \beta_j \gamma_k}= 
			\frac{\sqrt{2j+1}}{2\sqrt{1+\delta_{k0} \delta_{m0}}}
                                \frac{\cos( m \alpha_i+ k \gamma_k)}
                                {\sqrt{N_{\alpha} N_{\gamma}} }
                                \sqrt{w_{\beta_j}}d^j_{mk}(\cos \beta_j).
\end{equation}

\section{Kinetic Energy Matrix Elements in Water Dimer Calculations}

In this appendix, the matrix elements are given for the 
complex symmetric top basis
functions. Application of the formulas given here to the
 real symmetric top basis functions is straightforward.

If $\pbas= \vert j_1k_1m_1\rangle \vert j_2k_2m_2 \rangle$; then,
 the nonzero matrix elements of $\hat{K_1}$ is given by
\begin{equation}
\pbasd \hat{K_1}\pbas= \left(\frac{A+B}{2}\right)j_1(j_1+1) + 
                     \left(C- \frac{A+B}{2} \right)k_1^2,
\end{equation}
and 
\begin{eqnarray}
\pbasd \hat{K_1} \vert j_1k_1\pm2m_1j_2k_2m_2 \rangle & = &
          \left(  \frac{B-C}{4} \right)  \nonumber \\
    & & \times   \sqrt{j_1(j_1+1)-(k_1\pm2)(k_1\pm1)} \nonumber \\ 
              & &     \times      \sqrt{j_1(j_1+1)-(k_1\pm1)k_1}.
\end{eqnarray}

Similar expressions can be derived for $\hat{K_2}$. For $J=0$, nonzero
 matrix elements of $\hat{K_{12}}$ are given by
\begin{equation}
\pbasd \hat{K_{12}}\pbas  =  \frac{1}{2\mu R^2} 
                [j_1(j_1+1)+j_2(j_2+1)+ 2 m_1 m_2]
\delta_{m_1,-m_2},
\end{equation}
and 
\begin{eqnarray}
\langle j_1k_1m_1\mp1j_2k_2m_2\pm1 \vert \hat{K_{12}}\pbas &  = & 
          \frac{1}{2\mu R^2} 
     C^{\mp}_{j_1,m_1}  \nonumber \\ & & \times 
             C^{\pm}_{j_2, m_2}\delta_{m_1,-m_2},
\end{eqnarray}
where 
\begin{equation}
C_{jm}^{\pm}=\sqrt{j(j+1)\pm m(m+1)}.
\end{equation}

\newpage

\newpage
\bibliography{bibliography}

\end{document}